\documentclass[12pt]{iopart}
\usepackage{epsfig,psfig}
\usepackage{natbib}

\usepackage{color}
\newcommand{\be}{\begin{equation}}
\newcommand{\ee}{\end{equation}}
\def \cntl {\centerline}

\def\cf{{\it cf}\ }  %YH: new definition
\def \eg {{\it e.g.}\ }

\def  \LCDM{$\Lambda$CDM}

\newcommand{\hmsun}{{\,\rm h^{-1}M}_\odot}
\newcommand{\hmpc}{{\,\rm h^{-1}Mpc}}

\def \mnras {MNRAS}
\def \apj {ApJ}
\def \aj {AJ}
\def \aapr {A\& A Rev.}
\def \apjl {ApJL}
\def \apjs {ApJS}
\def  \aap {A\& A}
\def  \prd {Phys. Rev. D}
\def \physrep {Phys.Rep.}

\begin{document}

\title[The Future of the Local Large Scale Structure]{
The Future of the Local Large Scale Structure:
the roles of Dark Matter and Dark Energy\\
}

\author{Yehuda Hoffman$^1$, Ofer Lahav$^2$, Gustavo Yepes$^3$, and
  Yaniv Dover$^1$  }
\address{$^1$Racah Institute of Physics, 
Hebrew University, 
Jerusalem 91904, Israel }
\address{$^2$ Department of Physics and Astronomy, University College London,
Gower Street, London WC1E 6BT, UK}
\address{$^3$Grupo de Astrof\'{\i}sica, 
Universidad Aut\'onoma de Madrid,
Madrid E-280049, Spain }

\begin{abstract}

We study the distinct effects of Dark Matter and Dark Energy on the
future evolution of nearby large scale structures 
using constrained N-body simulations.  We contrast a model of Cold
Dark Matter and a Cosmological Constant (\LCDM) with an Open CDM
(OCDM) model with the same matter density $\Omega_m =0.3$ and the same
Hubble constant $h=0.7$.  Already by the time the scale factor
increased by a factor of 6 (29 Gyr from now in \LCDM; 78 Gyr from now
in OCDM)  the comoving position of the Local Group is frozen.  Well
before that epoch the two most massive members of the Local Group, the
Milky Way and Andromeda, will merge.  However, as the expansion rates
of the  scale  factor in the two models are different, the Local
Group will be receding in physical coordinates from Virgo
exponentially in a \LCDM\ model and at a  roughly  constant velocity in an OCDM
model.  More generally, in comoving coordinates the future large scale
structure will look like a sharpened image of the present structure:
the skeleton of the cosmic web will remain the same, but clusters will
be more `isolated' and the filaments will become thinner.  This
implies that the long-term fate of large scale structure as seen in
comoving coordinates is determined primarily by the matter density.
We conclude that although the \LCDM\ model is accelerating at present
due to its Dark Energy component while the OCDM model is non
accelerating, their large scale structure in the future will look very
similar in comoving coordinates.

\end{abstract}

\submitto{Acepted for publication in JCAP}

\maketitle

\section{Introduction}

 Cosmologists usually study the past evolution of the universe.
 Indeed, the Local Universe has been used before as a laboratory for
 studying dynamics and cosmological parameters, for example the timing
 argument 
  (\citeauthor{RayDLB89}  \citeyear{RayDLB89}, \citeauthor{kar05} \citeyear{kar05}), 
 or N-body methods
 \citep{PeebMel89}.  Similarly, we can gain insight into the dynamics
 and the role of the Dark Matter and Dark Energy components of the
 universe by considering the {\it future} evolution of nearby large
 scale structure.  The background cosmology is now reasonably well
 established based on the measurements of the cosmic microwave
 background, the large redshift surveys, Supernovae Ia data and other
 probes (e.g. \citeauthor{wmap3} \citeyear{wmap3} and references therein).  These
 measurements support a `concordance' model in which the universe is
 flat and contains approximately 4\% baryons, 21\% Cold Dark Matter
 and 75\% Dark Energy, with a possible small contribution of massive
 neutrinos.  The nature of the Dark Matter and the Dark Energy are
 still to be understood, in particular the possibility of a more
 general equation of state of the Dark Energy component, $w = P/\rho$,
 which might also be evolving with cosmic epoch.  The case $w=-1$
 corresponds to Einstein's Cosmological Constant $\Lambda$, which we
 shall consider in this paper, as it is consistent with the current
 observations.  We shall refer to this model as $\Lambda$ Cold Dark
 Matter (\LCDM).

 The fate of a  \LCDM\ universe as a whole has been discussed before
\citep[\eg][]{AL02,KrasSch07}.  
%OL_23Sep07
The goal of the present study is to contrast the \LCDM\ model with an OCDM model
with the same matter density parameter, to investigate the roles of
the Dark Matter and the Dark Energy in the evolution of clustering.
When the universe will be
 completely dominated by the $\Lambda$ term the scale factor will
 expand exponentially, $a(t) \propto \exp({\sqrt { \Lambda/3}}t)$ (a
 `de-Sitter phase').  It is less clear what will happen to the growth
 of structure and to individual objects in the universe. e.g. will
 they manage to survive as bound objects in an exponentially expanding
 universe?  Papers by 
 \citet[hereafter NL03]{NagLoeb03}
 and \citet{Busetal03}   have already considered some of these questions.

In particular NL03 used a constrained N-body simulation for a  \LCDM\ universe to
 predict the future evolution of nearby large scale structure.  They
 found that structures will freeze in comoving coordinates.  In
 particular they found that ``the Local Group will get somewhat closer to the
 Virgo cluster in comoving coordinates, but will be pulled away from
 Virgo in physical coordinates due to the accelerated expansion of the
 universe.'' 
 However, one should recall that the freeze out of the growth of structure is not the signature of the cosmological constant  but rather of the fact that the density of matter is less than the critical. Namely, in a universe dominated either by the negative curvature or by the $\Lambda$ term gravitational instability stalls. 
%OL_23Sep07
The linear theory growth of density perturbations in a universe 
with and without a cosmological constant is illustrated in numerous 
papers (e.g. Figure 2 of \citep{LS04}).
This has motivated us to explore the fate of the universe in the general case where the mean matter density is subcritical and compare the cases of a $\Lambda$ dominated universe with an open one.

Here we extend the NL03 study to explore to what extent this fate of
the local structure depends on the Dark Matter and Dark Energy contents of the
universe.  We use the tool of Constrained Realizations of Gaussian
random fields \citep{HR91} to constrain the initial
conditions for the N-body simulations by observational data.  The
resulting simulations reproduce quite faithfully the observed LSS out
to a few tens of Mpc from the LG.  We apply the simulations to both
\LCDM\   
model and an Open Cold Dark Matter (OCDM) model with the same
$\Omega_m=0.3$ and the same Hubble constant $h=0.7$.  We look in
particular at the universe at present epoch $a=1$ ( where the age of
the universe is 13.5 Gyr and 11.3 Gyr in  \LCDM\ and OCDM, respectively)
and at $a=6$ (where the universe will be 42.4 Gyr old for  \LCDM\ and
89.2 Gyr old in OCDM).  We find that in comoving coordinates the
evolution of large scale structure is quite similar.  The main
difference is in physical coordinates, as the evolution of the scale
factor is obviously different for the two models.

The outline of the paper is as follows.  In section 2 we summarize
briefly the properties of the Local Group and the Local Supercluster.
In Section 3 and in the Appendix we present analytic considerations
for the evolution of structure and in Section 4 we give details of the
constrained simulations.  The results and plots from the simulations
are shown in Section 5 and we discuss the results in Section 6.

\section{Nearby Structures in the Local Universe}

The Local Universe has been mapped in great detail since the 1980s, 
with the aid of whole sky galaxy surveys like IRAS, 2MASS
and peculiar velocity surveys (e.g. \citeauthor{sw95} \citeyear{sw95}  for a review).
Roughly speaking by Local Universe we mean the volume of  a sphere of radius
$\approx 200 h^{-1}$ Mpc centred at the Milky Way (MW).  
Here  we focus on the Local Group (LG) and the Local Supercluster (LSC).
For definitions of the local structure see e.g.   \citet{vdberg99}
and \citet{tf-atlas87}. 
Here we only  summarize briefly the terminology of local structures relevant for our study.
The Milky Way, Andromeda (M31) and other 30 other small galaxies within a few Mpc form the 
LG. At the present epoch the 
two major galaxies of the LG are approaching one another at an infall
velocity of about $120$ km/s and thus they constitute a bound dynamical
system. 
The distance between the Milky Way and M31 is $740 \pm 40$ kpc.
The total mass of the LG is estimated to be $2.3 \pm 0.6 \times 10^{12} M_\odot$ 
 \citep{vdberg99}.

Until the early 1980s the Virgo cluster was regarded as the centre of
an overdense region called the LSC 
(\eg,   \citeauthor{dh82}  \citeyear{dh82},  \citeauthor*{lyj86} \citeyear{lyj86})
 and it was assumed to be the major supercluster
in the local universe.  However, in the late 1980's it was recognized
in whole-sky surveys that much larger superclusters, such as the Great
Attractor and Perseus-Pisces, dominate the local universe. These and
other superclusters and voids generate tidal forces which affect the
motion of the LG towards the Virgo cluster.  Despite this complexity
of the local structure we find it useful below to consider the time
evolution of the distance between the LG and the centre of Virgo.
Obviously in principle we can define other distances or statistics to
quantify the local structure.
We also note the Supergalactic coordinate system,  defined based on a planar 
structure in the galaxy distribution \citep{devauc75},
which we shall use for convenience in some of our plots.

The Virgo-LG system is much less dynamically evolved than the LG itself.
 The observed mean overdensity in the number count of galaxies
within the Virgocentric sphere is $\approx 2$ 
%OL_23Sep07
\citep{dh82}
and the (line of sight)
peculiar velocity of the Virgo cluster relative to the LG is $932$
km/s at a distance of $16 \pm 2$ Mpc   \citep{vdberg99}.

Both the LG and the LSC constitute departure from the homogeneous and
isotropic expanding universe. An unperturbed open universe or a flat
$\Lambda$ dominated universe expand forever. Bound objects that have
collapsed and virialized by the present epoch will remain so in spite
of the future expansion of the universe. The question is what is the
fate of the objects in the nearby universe when the universe will be
freely or exponentially expanding. In particular we shall follow the
evolution of the two objects that dominate local
dynamics, namely the LG and LSC.

\section{Theoretical expectations}
\label{sec:th}

A rough estimate on the dynamical evolution is provided by the
spherical top-hat model \citep{gg72}.  The evolution of
spherical density perturbations in a general Friedmann universe
dominated by non-relativistic matter and a Cosmological Constant was
considered e.g.  by \citet{ol91}, \citet{ws98} and
\citet[hereafter LH01]{lh01}, \citet{mao05} and \citet{par05}.  A brief summary of the LH01 results
and an extension to include velocity perturbations as well are
presented in the Appendix.

Using the formalism presented in the Appendix the following values for the critical over-density to  future 
collapse are obtained. For the case of a vanishing  velocity
perturbation the critical over-density is 17.6 ($\Lambda$CDM; in agreement with NL03) and 2.33 (OCDM). Adopting a  Virgocentric infall 
velocity of $16 H_0- 932=188 km/s$  
(for $H_0 = 70 km/sec/Mpc$)
with  the critical value  is 
14.6 ($\Lambda$CDM) and 1.3 (OCDM)
(\cf  Appendix).
Given the observed Virgocentric overdensity (in
galaxy count) of $\approx 2$ one can safely assume that in the
$\Lambda$CDM model the Virgocentric infall is expected not to proceed
to a collapse and virialization but rather to reach a freeze out. The
case of the OCDM is not clear as the current estimation of the
Virgocentric infall and overdensity are only marginally consistent
with a freeze out of the infall. Numerical simulations are needed here
to resolve the issue of the future of the LSC in the OCDM model.

The LG has already passed its turn-around phase and the MW and M31 are
heading at a (line of sight) velocity of 120 km/s towards a merger
to become one object. The mean overdensity in the LG is $\approx 260$
(by averaging the mass of the MW 
and M31 
over a sphere with a radius which is 
half the distance MW-M31)
and it is much larger than $\Delta_{crit}$ for both the $\Lambda$CDM
and OCDM models. It follows that we expect the LG to collapse and
merge into one object in both cosmologies.

\section{N-body simulations: methodology}

 The goal of the present paper is to study the future evolution of the
 nearby LSS in both OCDM and a flat \LCDM\  cosmologies.
The  nearby universe seems to constitute a
 very typical realization of the flat-$\Lambda$CDM and the OCDM
 models. The LG, in particular, is not a unique or an unusual object
 in the universe, yet not being a relaxed object in virial equilibrium
 it has its own characteristics that affect the outcome of any
 dynamical test that would be applied to it.  The key to a successful
 numerical study of the LG and the nearby universe is the ability to
 reproduce the characteristic dynamical properties of the LG  and its surrounding.  
 Namely a
 quasi-linear object located at about 
  $(10-15)\hmpc$  
 from a Virgo-like
 cluster, within a supercluster that extends as filament connecting
 two major structures, the Perseus-Pices supercluster and the Great
 Attractor that are located some $90\hmpc$ apart. The LG is caught in
 a `tug of war' in between these two major structures, which affects
 the local dynamics. The key for a successful numerical study of the
 local universe is the ability to design numerical simulations which
 reproduce the main dynamical features of the local universe. The
 optimal way of achieving that goal is by the use of constrained
 simulations (CSs; \citeauthor{kkh02}  \citeyear{kkh02}, 
 \citeauthor{gif02} \citeyear{gif02},  
 \citeauthor{khkg03} \citeyear{khkg03}), 
 namely simulations based on initial conditions set by means of
 constrained realizations of Gaussian fields \citep{HR91}.

The data used to constrain the initial conditions of the simulations is
made of two kinds. The first data set is made of radial velocities of
galaxies drawn from the MARK III \citep{mark3}, SBF \citep{sbf01} and
the \cite{kar05} catalogs. Peculiar velocities are less affected by
non-linear effects and are used as constraints as if they were linear
quantities \citep{zhd99}. This follows the CSs
performed by \cite{kkh02} and 
\cite{khkg03}.  
The other constraints are
obtained from the catalog of nearby X-ray selected clusters of galaxies
\citep{rei02}. Given the virial parameters of a cluster and assuming
the spherical top-hat model one can derive the linear overdensity of
the cluster. The estimated linear overdensity is imposed on the mass
scale of the cluster as a constraint. 
It should be noted that neither the MW and M31 nor the LG have been imposed directly on the simulations by the constraints. The constraints used here constrain quite closely the structure on scale larger than $\approx 5\hmpc$  \citep{khkg03}. 
Different CSs with different
random realizations have been calculated and they all exhibit a clear
and unambiguous LSC-like structure that dominates the entire
simulation, much in the same way as in the actual universe in which the
LSC dominates the nearby LSS. The lack of explicit LG constraints causes the simulations to vary with respect to
the particular details of the LG-like object. Yet, the fact that the large scale structure of the local universe is faithfully reproduced  implies that the CSs have a high probability of producing quite reasonable LG-like objects. This is definitely confirmed by the present simulations.
%YH

%LCDM
We assume a flat  \LCDM\ model with
$\Omega_m=0.3$, $\sigma_8=0.9$ and $h=0.7$ (where $\sigma_8$ is the
power spectrum normalization factor).
A somewhat refined set
of cosmological parameters has been adopted after WMAP third year data
release  \citep{wmap3}, but the change in the values of the
cosmological parameters would not 
affect the results presented here in any significant way. 
%OCDM
The other model used here is the OCDM, which   is obviously inconsistent with recent 
observations of the CMB  \citep{wmap3}. 
The model is considered here so as to show that the main affect that drives the freeze out of structure is the fact that $\Omega_m$ is less than unity. For the sake of concreteness we assume that the cosmological parameters of the OCDM model coincide with that of the  \LCDM\ model apart from setting $\Omega_\Lambda=0$, and thereby implying an open universe.
The simulations corresponds to a periodic  cubic box of  64 h$^{-1}$ Mpc  on a
side, spanned by a $256^3$ grid.
This translates into a mass per
particle of 
$1.3 \times  10^9 h^{-1}  M_\odot$. 
The softening of the gravitational force  corresponds to an  equivalent Plummer softening of  comoving 5 kpc/h, which reproduce the  exact Newtonian force at 2.8 times this scale, implying a 
 force resolution of    15 kpc/h.
The mass resolution is adequate for resolving a LG-like object made of two $\approx 10^{12} \hmsun$ objects and following its internal dynamics. Yet, the internal dynamics of the main halos cannot be resolved. This to be compared with the mass resolution  of $3.6\times 10^{11}\hmsun$ of NL03, which does not resolve the LG.
The current simulations are among the ones used by 
\citet{mvyh07}. The parallel TREEPM N-body code  GADGET2 
(\cite{gadget2}) has been used to run the  simulation.
A more detailed description of the simulation is presented in \citet{mvyh07}. The halos are found by the AMIGA \citep{amiga} halo finder, and the mass $M$ corresponds to the mass enclosed within the virial radius.

\section{Constrained simulation of the local universe}

The CS clearly manifests the main characteristics of the local universe. Fig. \ref{fig:whole} shows the projected dark matter density of a $12\hmpc$ thick slice centered on the Supergalactic Plain in both the 
$\Lambda$CDM and OCDM models. The local structure is dominated by the LSC filaments which crosses horizontally the Supergalactic Plain at roughly $SGY \approx 15 \hmpc$. The simulated LG is located in a filaments that run perpendicularly to the LSC at $SGX \approx -7\hmpc$. Apart from a general shift by a few Mpc in the $-SGX$ direction of the of the whole cosmic web, the simulated structure recovers the observed one. It should be noted at the outset that the present (and future)  LSS exhibited by the two models looks very similar. 
The main dynamical parameters that characterize the simulated local universe at the present and future epochs for both the $\Lambda$CDM and OCDM models are presented in Table  1.
Both models reproduce the structure of the LSC at the present epoch with a simulated LG located 13.0 ($\Lambda$CDM) and 11.6 Mpc/h (OCDM) away from the LG, compared with the actual value of 11.2Mpc/h. The mass of the simulated Virgo 
is $1.7\ 10^{14}\hmsun$ 
($\Lambda$CDM) and $0.7\ 10^{14}\hmsun$ (OCDM), compared with the observed $\approx 10^{14}\hmsun$.
The mass of the simulated LG is very close to the observationally inferred value of $1.6\ 10^{12}\hmsun$   \citep{vdberg99}. The OCDM simulated MW-M31 distance is close to the observed value of 0.7 Mpc, but the $\Lambda$CDM simulated distance is more than twice larger. That last discrepancy does not affect in any way the conclusions drawn here on the future evolution of the  local universe.

\begin{table}[htdp]
\begin{center}
\begin{tabular}{ccccc}
\hline \hline
Model/epoch & $\Lambda$CDM\ a=1 &  OCDM\ a=1 & $\Lambda$CDM\ a=6  & OCDM a=6 \\ 
 \hline
 $M_{MW}$ [$\hmsun$]    & $7.9\times 10^{11}$  & $5.5\times 10^{11}$   & \  & \                           \\ 
 \hline
 $M_{M31}$ [$\hmsun$]   & $1.0\times 10^{12}$  & $1.5\times 10^{12} $  & \  & \                           \\ 
 \hline
  $M_{LG}$ [$\hmsun$]     & $1.8\times 10^{12}$  & $2.0\times 10^{12}$  & $5.7\times 10^{12}$   & $2.4\times 10^{12}$    \\ 
 \hline
  $R_{LG}$ [$\hmpc$]        & 1.1         & 0.44      & \  & \                              \\ 
  \hline
 $M_{Virgo}$ [$\hmsun$] & $1.7\times 10^{14}$ & $6.8\times 10^{13}$  & $3.4\times 10^{14}$   &  $5.6\times 10^{14}$     \\ 
 \hline
  $R_{LSC}$ [$\hmpc$]      & 13.0     & 11.6       & 10.6        & 15.6          \\ 
\hline \hline
\end{tabular} 
\caption{
The main dynamical parameters that characterize the simulated local universe: mass of Milky Way ($M_{MW}$), mass of M31 ($M_{M31}$), mass of LG ($M_{LG}$, for the present epoch it is the sum of 
$M_{MW}$ and $M_{M31}$), mass of the Virgo cluster ($M_{Virgo}$) and the LG-Virgo distance ($R_{LSC}$). Here,  MW, M31 and the Virgo cluster refer to the simulated objects.
}
\end{center}
\label{table1}
\end{table}

%YH

\subsection{The evolution of the LSC}

\begin{figure*}
\begin{center}
\vskip -5.75cm
\cntl{ \includegraphics[scale=1.0]{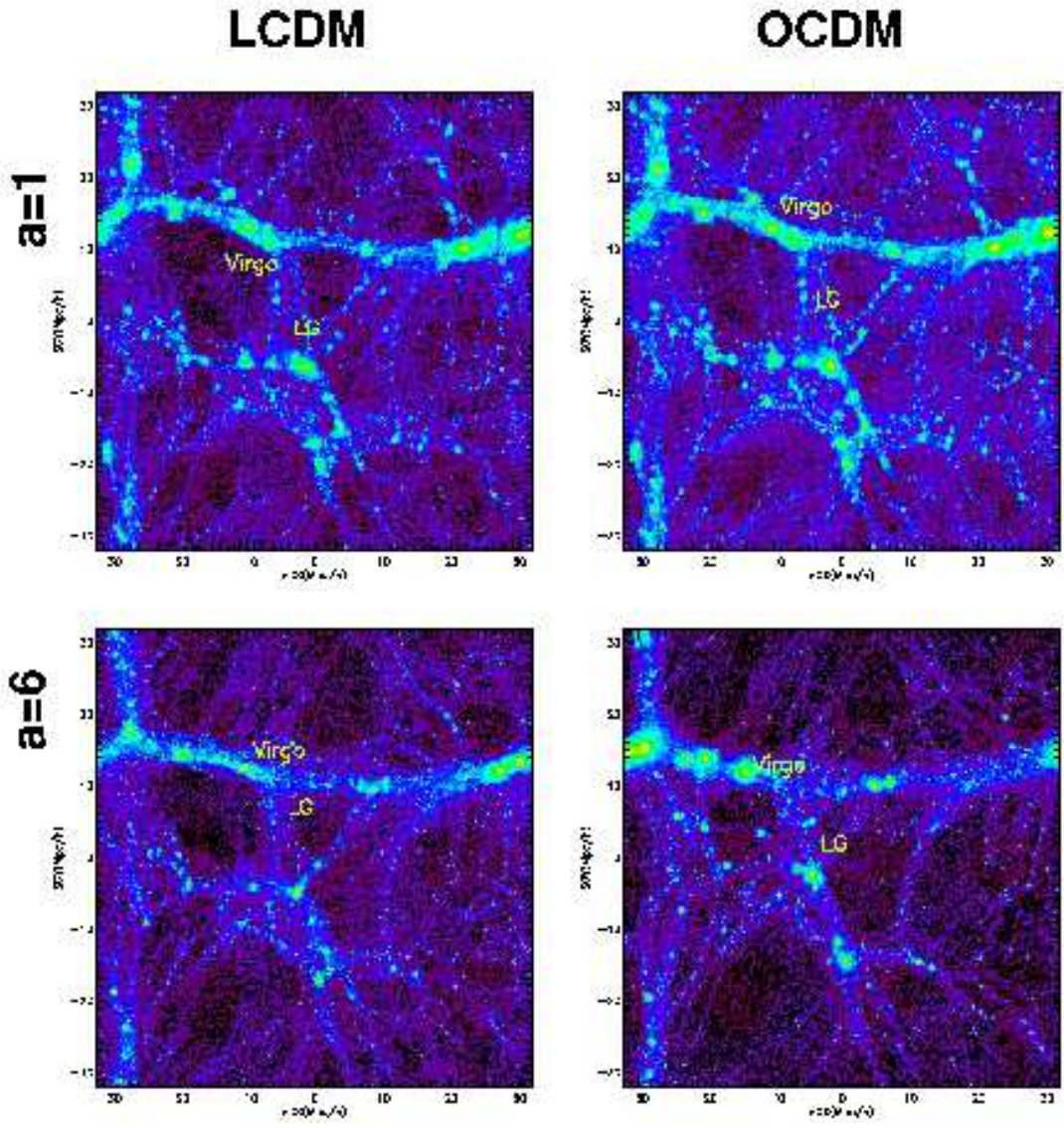} }
%\cntl{ \includegraphics[scale=1.1]{HLYD_Fig1.ps} }
\vskip -7.5cm
\end{center}
\caption{
The mass distribution in the Supergalactic plane in the $\Lambda$CDM (left column) and OCDM (right column) models presented  at the present epoch ($a=1$; upper row) and the future ($a=6$; lower row). 
The plot shows the projection of the particles in a slab of thickness $12 \hmpc$, centered on the Supergalactic plane. The color coding of the particles indicate the density field. The positions of the LG and the Virgo cluster are marked.
}
\label{fig:whole}
\end{figure*}

The $\Lambda$CDM simulated local universe is presented at the present epoch of $a=1$ 
and  future epoch of $a=6$ (Fig. \ref{fig:whole}), where $a$ is the expansion factor. The projected density maps are presented in term of co-moving Supergalactic coordinates. An inspections of 
Fig.   \ref{fig:whole} reveals that the cosmic web at the two epochs has hardly changed. Yet, the late epoch web is more `skinny' and is more dominated by massive halos and less by the smaller mass halos.

\begin{figure*}
\begin{center}
\resizebox{\linewidth}{!}{\includegraphics{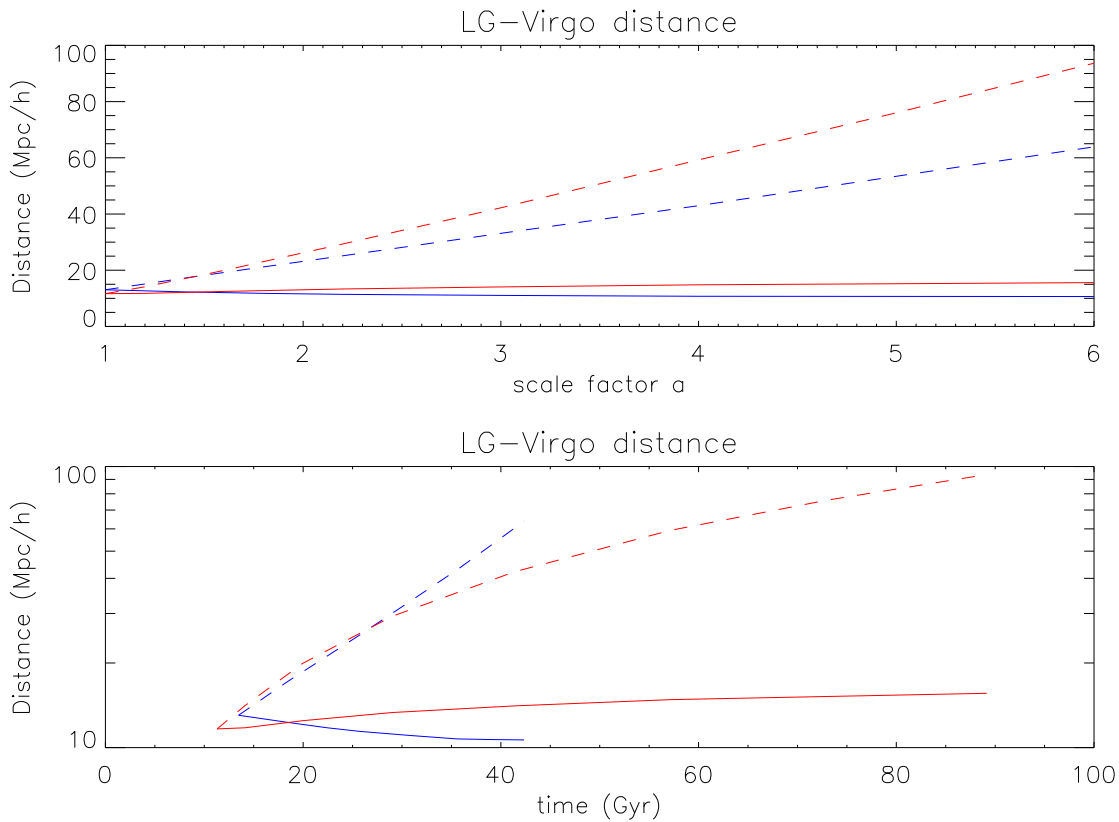}}
\end{center}
\caption{
The evolution of the distance of the LG from the Virgo cluster in the $\Lambda$CDM (blue curves) and 
OCDM (red curves) models is presented in comoving (solid line) and physical (dashed lines) coordinates. The upper panel presents the distance as a function of the scale factor $a$ and the lower  one as a function of time.
In the $\Lambda$CDM case   the Virgocentric infall extends to about twice the present Hubble time before it converges to its asymptotic value, where the inflow freezes. In physical coordinates  the distance increases at all time and eventually it grows exponentially. In the OCDM the time evolution of the distance does not show the freezing of the flow. The position of the LG indeed becomes frozen in comoving coordinates but the Virgo cluster  moves away from it because of the internal dynamics within the  LSC. 
}
\label{fig:LGVdist}
\end{figure*}

It should be noted here that the similarity between the present and future structure exists only upon using co-moving coordinates. In physical coordinates the cosmic web gets more empty and the typical length scale of the web expands exponentially. This is clearly seen in Figs. \ref{fig:LGVdist}   which shows  the evolution of the the Virgo-LG distance (in physical and co-moving coordinates). The first 30 Gyrs  reflect the Virgocentric infall of the LG towards the Virgo. The infall is manifested in the co-moving sense only and it levels off in about 35 Gyrs. In physical coordinates the LG is receding away from the Virgo cluster almost exponentially. Note that   after $t_0 +  t_H$  the comoving distance remains constant with time, where $t_H=H^{-1}=\sqrt{\Lambda/3} \sim 1.7H{_0^{-1}}$, in  agreement with  \citet{NagLoeb03}.

The present ($a=1$) and future ($a=6$) simulated local structure in the OCDM model are displayed in Fig. \ref{fig:whole}. The LSS, viewed in comoving coordinates exhibited by the $\Lambda$CDM and OCDM models at the two epochs shows a great resemblance. The general appearance of the cosmic web of the two models is almost indistinguishable. Yet, a close inspection of the LSC shows some differences. The LSC almost freezes  out in the   $\Lambda$CDM model. In the OCDM case the center of the LSC keeps on evolving. In particular the Virgo cluster moves along the filament that constitutes the LSC, from right to left in the Supergalactic projection, and merges with other clusters. The LG, on the other hand, is essentially frozen in comoving coordinates. As a result the Virgocentric distance of the LG keeps on growing in comoving coordinates.

\subsection{The evolution of the LG}

\begin{figure*}
\begin{center}
\vskip -4.25cm
\cntl{ \includegraphics[scale=0.8]{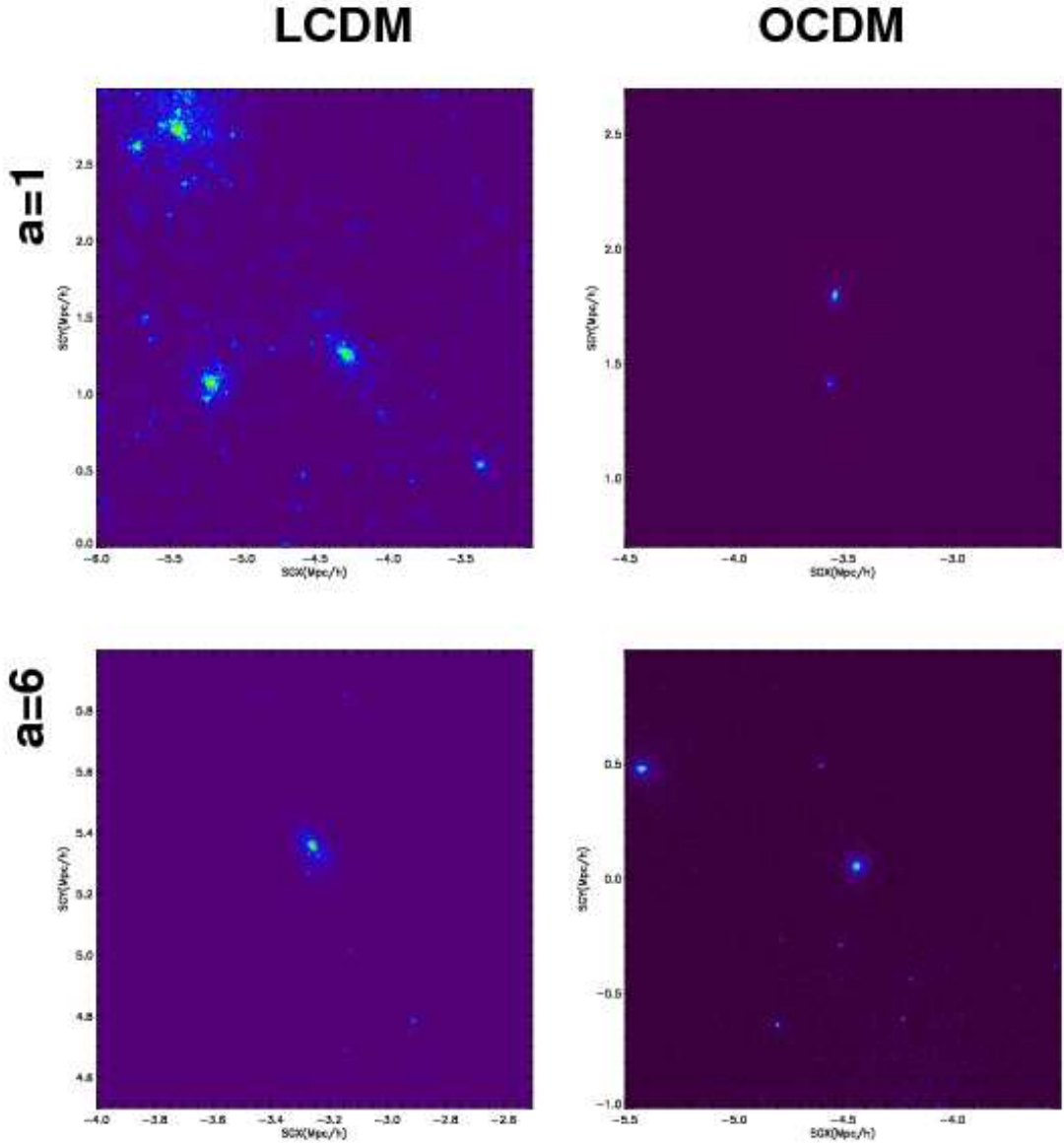} }
\vskip -5.cm
\end{center}
\caption{
The mass distribution around the LG  in the Supergalactic plane in the $\Lambda$CDM (left column) and OCDM (right column) models presented  at the present epoch ($a=1$; upper row) and the future ($a=6$; lower row). 
The plot shows the projection of the particles in a slab of thickness $3 \hmpc$. The color coding of the particles indicate the density field.  At the present epoch the LG is dominated by two massive halos of mass $\approx 10^{12}\hmsun$
 at a distance of $\approx 1.1 \hmpc$ and $0.5 \hmpc$ in \LCDM\  and OCDM respectively. There is very little difference between the dynamics of the  
simulated LG in the   \LCDM\ and OCDM models.
}
\label{fig:LG}
\end{figure*}

As the observed distance between the Milky Way and Andromeda is 740
kpc and they are approaching one another at 120 km/sec, we can
estimate naively that they will collide in about 6 Gyr 
if relative transverse velocity is ignored \citep{cox07}. 
The simulated $\Lambda$CDM LG is
shown at the present epoch in Fig. \ref{fig:LG} exhibits two DM halos
with masses of about $10^{12}\hmsun$, at a distance $~1.1\hmpc$ and
located in a slightly overdense filament that connects the LG group
with the Virgo cluster of a mass of $~10^{14}\hmsun$ located
$~10\hmpc$ away.  The relative infall velocity of the two halos is
about 140 km/sec and the transversal relative velocity is 50 km/s. It
follows that the CS has reproduced a LG-like object that resembles the
actual LG apart from the MW-M31 distance which is  larger than the true
distance of $~0.74 Mpc$.  From the $\Lambda$CDM simulation we find
that the simulated MW and M31 merge into one object, at the epoch of
$a=3.0$, corresponding for $h=0.7$ to 
17.3 Gyr from
now (we note this is longer than that expected from just the two-body
radial motion based on the the simulation values: with infall of 140
km/sec starting at relative distance of 1570 kpc, we expect 11.0 Gyr).

Our large simulated present epoch MW-M31 distance obviously implies
that the estimated time till the LG merger is overestimated.  Indeed,
\cite{cox07} have recently estimated that the merger will take place
in $~5 Gyr$.

The simulated OCDM LG is very similar to the $\Lambda$CDM one, both in
structure and evolution (Fig. \ref{fig:LG}).  The present epoch MW-M31
distance of the OCDM LG is somewhat smaller than in the $\Lambda$CDM
case, $\approx 0.5\hmpc$ and therefore the MW-M31 merger is taking
place earlier, at $a \sim 1.1$, corresponding to (12.7-11.3) Gyr = 1.4
Gyr from now.
Fig. \ref{fig:LG}
shows the final merged object much later (at $a=6$) in both simulations.  

It is clear that in the present CSs the internal structure of the LG is very weakly constrained and different realizations exhibit a large scatter with respect to the structure of the LG. This leads to a considerable scatter in the collapse time of the LG. Yet, all simulated LG-like objects will collapse much before $a=1$.

\subsection{General remarks on the future evolution}

The cumulative mass function $n(>M)$ of DM halos serves as a  good indicator of the growth of structure of a cosmological model, where $n(>M)$ is the number of DM halos more massive than $M$ per unit co-moving volume. We have calculated the mass function of both the $\Lambda$CDM and OCDM models in the  present $a=1$ and future $a=6$ epochs. Generally speaking the mass function of the two models are very similar at both epochs. This similarity follows from the very similar dynamical nature of the two models, when analyzed in co-moving coordinates. The cumulative mass function hardly evolves in both models, yet at $a=6$ there are somewhat fewer small, $\approx 10^{12}\hmsun$,  DM halos and roughly twice as many rich cluster -like halos. A detailed analysis shows that the depletion in the number of  small halos and the appearance of more massive clusters is somewhat more pronounced in the OCDM model compared with the $\Lambda$CDM one.

A close inspection of Fig \ref{fig:LGVdist} shows that the LG-Virgo distance is already frozen by $a=6$ in the  $\Lambda$CDM model, while it is still slowly increasing in the OCDM case at that epoch. This is consistent with the evolution displayed by the mass function, namely by $a=6$ the OCDM model is somewhat more evolved than the $\Lambda$CDM model. The slight differences between the models is clearly understood.

A simple analytical reasoning leads to the understanding that once the dynamics of the universe is no longer dominated by the cold matter density, gravitational instability  comes to an end and the growth of structure freezes out. This property is shared by both the $\Lambda$CDM and OCDM models. Yet, the rate of transition to the asymptotic state is different. In the $\Lambda$CDM case the asymptotic  state is achieved when the (constant for $w=-1$) dark energy term  dominates the cold matter term (which decays like $a^{-3}$) in the Friedmann equation. In the OCDM case, on the other hand, the  freeze out is achieved when the curvature term (which decays as $a^{-2}$) dominates. It follows that the $\Lambda$CDM model converged much faster, in time and scale factor as ell,  to the freeze out state than the OCDM model.

\begin{figure*}
\begin{center}
%\vskip -4.25cm
\cntl{ \includegraphics[scale=0.6]{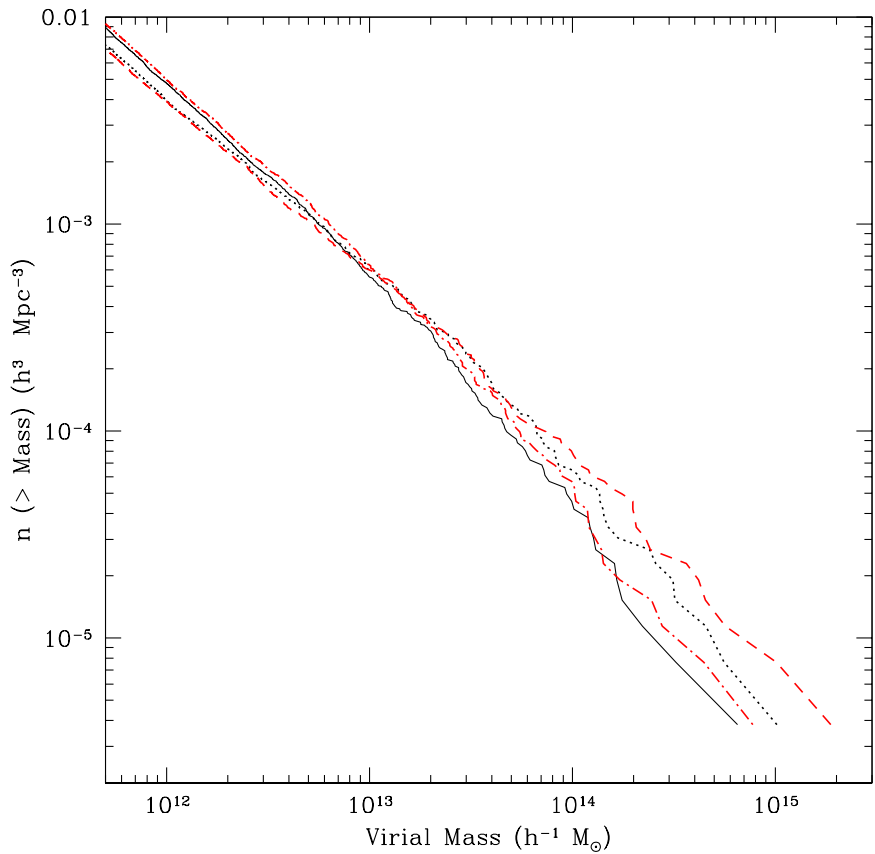} }
\vskip -1.5cm
\end{center}
\caption{
The cumulative mass function $n>M$ is plotted for both models and epochs,  $\Lambda$CDM $a=1$ (solid line, black) and $a=6$ (dotted, black) and for  OCDM $a=1$ (dot-dashed, red) and $a=1$ (dashed, red). DM halos are defined by the AMIGA halo finder and $M$  corresponds to the mass within the virial radius. The cumulative mass function $n(M)$ gives the number of DM halos per unit volumes with mass greater than $M$.
}
\label{fig:massfunc}
\end{figure*}

\section{Discussion}

Exploring the future of evolution of the large scale structure of the
local universe is more than just a curiosity.  \citet{Busetal03} have
demonstrated that by running simulations into the future it is
possible to identify more efficiently which objects will form bound
objects.  It also allows us to get further insight into the effect of
Dark Matter and Dark Energy in the evolution of large scale structure.
In particular we extended here the study of 
NL03 who analyzed with the aid of constrained N-body simulations the
evolution of nearby large scale structure only in a flat \LCDM\  model.
By contrasting the \LCDM\  simulation with an OCDM with the same matter
density $\Omega_m =0.3$ and the same Hubble constant $h=0.7$ we can
learn what is the role of the Dark Matter compared with the role of
the Dark  Energy. 

Our main conclusion is that 
the long-term fate of large scale structure as seen in comoving
coordinates is determined primarily by the matter density.  
This generalizes the result of NL03, and clarifies the distinct role
of the matter density in defining the freezing out of structure,
and the combined effects of both the Dark Matter and Dark Energy in the evolution 
of the scale factor and hence in the physical coordinates.

In more detail our main results are:

(i) The Milky Way and Andromeda will merge already at $a=3$ and at
$a=1.1$ according to our \LCDM\  and OCDM simulations (where the present
epoch distances between the two galaxies in our simulations are
somewhat different from the true observed value of 740 kpc). 

(ii) Already by $a=6$ the comoving position of the  LG becomes frozen at its asymptotic value.
In the  \LCDM\  case the comoving LG-Virgo distance reaches its asymptotic value. In the OCDM case the Virgo cluster still moves within the LSC and hence that distance is still slightly growing.   However, in physical coordinates the LG will be receding from Virgo exponentially in a \LCDM\
model and at an almost  constant velocity in an OCDM model.
This is just  a manifestation of the different evolution of the scale factor $a(t)$ in the two models.

(iii) When we consider the overall cosmic web we find that 
qualitatively the future large scale structure will look like a sharpened
image of the present structure: the skeleton of the cosmic web will
remain the same, but clusters will be more `isolated' and the
filaments will become thinner.

The discussion has  focused so far on the dynamical evolution of our cosmological neighborhood, which hardly distinguishes between the comoving structure of the  \LCDM\ and OCDM models. 
Yet, the two models appear fundamentally very different when viewed in physical coordinates and in particular when the notion of the horizon is considered. 
The comoving event horizon in the  \LCDM\ model of the present epoch at the present time is $3.4 h^{-1}Gpc$ and  at $a=6$ it is $0.6  h^{-1}Gpc$. 
As was pointed by  \citet{AL02} the 
 future LG observer  in the  \LCDM\ model will be living in an Island Universe. 
 This will happen at $a\approx350$ (namely, at $t \approx 8 H{_0^{-1}}$), when the Virgo cluster will be at the event horizon of the LG observer.
 The cosmic web will not be accessible  to these observers and the whole notion of comoving coordinates will be meaningless  \citep{KrasSch07}.

The work presented here can be extended further in a number of ways.
The evolution of the skeleton structure with time in different models
could be quantified by advanced statistics (\eg\   the probability
distribution function and Minkowski functionals).  The models could be
extended to study the much discussed equation  of  
state parameter $w$,
and to allow it to be epoch dependent, 
\eg\ $w(a) \approx  w_0 + w_a (1-a)$.  
Major surveys of Dark energy are underway to quantify $w(a)$ (\eg\  the
DETF report   \citep{de-report06}
and ESO/ESA report \citep{eso-esa-report06}, 
and it would be interesting to understand
the implications for the growth of structure in the universe if $w(a)$
turns out to be different than the $w=-1$ assumed in this paper.  In
addition, one may explore the impact of changes in the value of
$\Omega_m$, the effect of massive neutrinos (which suppress structure
on large scales) and the variations of the normalization of power spectrum 
$\sigma_8$.

Finally, we remark on the connection between our simulation results
and the Anthropic principle.  The basic Anthropic argument is that if
the matter density was too high the universe would have collapsed by
the present epoch without providing enough time for life to emerge and
evolve. On the other hand, if the matter density was too small,
structure would have not collapsed and again life would not have evolved.  It has been argued further that the observed value of the
cosmological constant can also be supported by Anthropic arguments,
(\eg\   \citeauthor{Wein97}   \citeyear{Wein97},  \citeauthor{Efst05} \citeyear{Efst05}, 
\citeauthor{TegRees98}  \citeyear{TegRees98}, \citeauthor{Wein05} \citeyear{Wein05}
and references therein). In a nutshell, Weinberg's original
argument was that the Cosmological Constant (in the form of a vacuum
energy) cannot be too large and positive, because then galaxies would
not form. The vacuum density should be less than the matter density of
the universe at the epoch when galaxies formed.  The probability of a
random observer seeing the value of the vacuum energy as small as
observed is about 15\%, under certain assumptions \citep{Wein05}.
We note in particular 
the sensitivity of the argument to the assumed amplitude of density fluctuations \citep{TegRees98}. More recently, the possibility of a \LCDM\  universe has been considered
in the context of a `string landscape', where the universe we live
in is only one of many in 
the `multiverse' 
(\citep{Wein05} and references therein).  It has been argued 
in the above references
that a
\LCDM\ universe is suitable for the emergence of life, while other
universes with other extreme values of matter and vacuum densities
would be hostile to life.

While the observed value of the Cosmological Constant is consistent
with this argument, we argue that as the large scale structures expected
in  \LCDM\   and the OCDM scenarios look so similar at the present epoch
($a=1$), there is no preference from anthropic arguments alone for one
or the other. The  dominant  
effect of the formation of structure (as
seen in comoving coordinates) is the matter density rather than the
vacuum energy.

\ack

Fruitful discussions with Avi Loeb are gratefully acknowledged.
We  thank CIEMAT (Spain)
for the use of the 
SGI-ALTIX supercomputer  and to NIC J\"ulich
(Germany) for the  access to the IBM-Regatta p690+ JUMP supercomputer.
GY would like to thank also MEC  
for financial support under 
 project numbers  FPA2006-01105 and AYA2006-15492-C03.
YH acknowledges the support of  ISF-143/02, the Sheinborn
Foundation and a PPARC Visiting Fellowship at UCL. 
OL acknowledges a PPARC Research Senior Fellowship.

\section*{Appendix}

The fate of an individual object in an expanding universe can be readily calculated under the assumption of spherical symmetry. Here we follow the analysis of LH01  for the evolution of density perturbations in a universe made of (non-relativistic) matter and gravitational constant of an arbitrary curvature. The LH01 analysis is repeated here for the sake of completeness and it is simply extended to include velocity as well as density perturbations.

The dynamical analysis of spherical perturbations often starts with the Tolman-Bondi energy-like term  of a spherical shell of radius $r$,
\begin{equation}
\label{eq:TB}
\epsilon   = {v^2 \over 2} - {G M (r)\over r} -  {4 \pi \over 3} G \rho_\Lambda r^2,
\end{equation}
where $M(r)$ is the mass enclosed within a radius $r$ and $\rho_\Lambda={\Lambda c^2\over 8 \pi G}$.
In the absence of shell crossing $\epsilon$ is a constant of motion.
This can be rewritten as:
\begin{equation}
\label{eq:energy}
\epsilon   =\Big(  \big({v \over v_H}\big)^2 
                       -  \Omega_m (1+\Delta)  
                        -  \Omega_\Lambda  \Big)  {H{_0^2} r^2 \over 2},  
\end{equation}
where $\Delta$ is the cumulative overdensity within the radius $r$, $v_H=H r$ and $H$ is Hubble's constant. It follows that the integral of motion can be evaluated at any time, denoted here as $t_i$. Thus the R.H.S. of Eq. \ref{eq:energy} is evaluated at $t_i$ when the cosmological parameters equal $\Omega_{m,i}$, $\Omega_{\Lambda,i}$ and $H_i$ and the density perturbation is $\Delta_i$. For convenience $t_i$ is assumed to precede the time of turn-around of the given shell.

The equation of motion of the radius of the shell is readily given by 
\begin{equation}
\label{eq:eom}
    \frac{{\rm d} s}{{\rm d} t} =H_{\rm i} \left[ 1 + \Omega_{m,\rm i} (\Delta_{\rm i}+1)
    \left( \frac{1}{s} - 1 \right) + \Omega_{\Lambda, \rm i} (s^2 -1)
    \right]^{1/2},
\end{equation}
where $s$ is the radius scaled by its value at $t_i$, $s=r/r_i$. 
The turn-around radius is found by setting the R.H.S. of  \ref{eq:eom} to zero. The condition for a turn-around radius to exist is that $\Delta_{\rm i}$ should be larger than a critical density given by:
\begin{equation}    
\label{eq:del-cr}
 \Delta_{\rm  cr, i} = \frac{1}{\Omega_{\rm i}}
    p(\Omega_{\Lambda,\rm i}) - 1
\end{equation}
where
\begin{equation}    
\label{eq:del-cr-a}
    p(\Omega_{\Lambda,\rm i}) = 1 + \frac{5 \Omega_{\Lambda,\rm i}}{4}  +
    \frac{3 \Omega_{\Lambda,\rm i} (8 + \Omega_{\Lambda,\rm i})}{4
   q(\Omega_{\Lambda,\rm i})} + \frac{3 q(\Omega_{\Lambda,\rm i})}{4}
\end{equation}
and
\begin{equation}    
\label{eq:del-cr-b}
    q(\Omega_{\Lambda,\rm i}) = \{ \Omega_{\Lambda,\rm i} [8 - \Omega_{\Lambda,\rm i}^2 + 20
    \Omega_{\Lambda,\rm i} + 8 (1-\Omega_{\Lambda,\rm i})^{3/2}] \}^{1/3}
\end{equation}
%YH:
The LH01 solution (Eqs. \ref{eq:del-cr} - \ref{eq:del-cr-b}) is obtained for the case of no velocity perturbation, namely $v/v_H=1$.

The  LH01
analysis can be easily extended to accommodate also velocity perturbations.
This is done by expressing the velocity perturbation as a change of the global Hubble constant into a local one. Consider a spherical shell at the fiducial time $t_i$ with a peculiar velocity  $v_{\rm p,i}$. The Tolman-Bondi energy equation (Eq. \ref{eq:energy})is rewritten as,

\begin{equation}
\label{eq:energy-i}
\epsilon   =\Big(  \big(1+{v_p \over v_H}\big)^2 
                       -  \Omega_m (1+\Delta)  
                        -  \Omega_\Lambda  \Big)  {H{_0^2} r^2 \over 2},  
\end{equation}
and here a positive $v_p$ indicates an outflow. An  effective local Hubble constant is defined by by
\begin{equation}
\label{eq:hl}
h_l=H (1+{v_p \over v_H})
\end{equation}
and effective local density parameters by
\begin{equation}
\label{eq:omegal}
\omega_x= \Omega_x (1+{v_p \over v_H})^{-2},
\end{equation}
where $x$ stands form $m$ or $\Lambda$.

Eq. \ref{eq:eom} can now be easily solved by replacing the global cosmological parameters $H$, $\Omega_m$ and $\Omega_\Lambda$ by the local values of $h_l$, $\omega_m$ and $\omega_\Lambda$ evaluated at $t_i$. 

Eq. \ref{eq:del-cr}  implies that in the case of no velocity perturbation the critical overdensity is 17.6 ($\Lambda$CDM) and 2.33 (OCDM). In the case of velocity perturbations Eq. \ref{eq:del-cr} is still valid, but with a modified local Hubble constant and density parameter (Eqs. \ref{eq:hl} and \ref{eq:omegal}).
Adopting for the Virgo cluster a distance of 16Mpc, a Hubble constant of $H_0=70 km\ s^{-1}\ Mpc^{-1}$ and a radial recession velocity of 932km/s (\cf\ \S\ref{sec:th}), one finds $(1+v_p/v_H)=0.83$.
It follows that
the critical overdensity for the Virgocentric infall is 14.6 ($\Lambda$CDM) and 1.3 (OCDM).

\References

% \begin{thebibliography}{38}
\expandafter\ifx\csname natexlab\endcsname\relax\def\natexlab#1{#1}\fi

\bibitem[{{Albrecht} {et~al.}(2006){Albrecht}, {Bernstein}, {Cahn}, {Freedman},
  {Hewitt}, {Hu}, {Huth}, {Kamionkowski}, {Kolb}, {Knox}, {Mather}, {Staggs},
  \& {Suntzeff}}]{de-report06}
{Albrecht}, A., {Bernstein}, G., {Cahn}, R., {Freedman}, W.~L., {Hewitt}, J.,
  {Hu}, W., {Huth}, J., {Kamionkowski}, M., {Kolb}, E.~W., {Knox}, L.,
  {Mather}, J.~C., {Staggs}, S., \& {Suntzeff}, N.~B. 2006, ArXiv Astrophysics
  e-prints

\bibitem[{{Busha} {et~al.}(2005){Busha}, {Evrard}, {Adams}, \&
  {Wechsler}}]{Busetal03}
{Busha}, M.~T., {Evrard}, A.~E., {Adams}, F.~C., \& {Wechsler}, R.~H. 2005,
  \mnras, 363, L11

\bibitem[{{Cox} \& {Loeb}(2007)}]{cox07}
{Cox}, T.~J. \& {Loeb}, A. 2007, ArXiv e-prints, 705

\bibitem[{{Davis} \& {Huchra}(1982)}]{dh82}
{Davis}, M. \& {Huchra}, J. 1982, \apj, 254, 437

\bibitem[{{de Vaucouleurs}(1975)}]{devauc75}
{de Vaucouleurs}, G. 1975, \apj, 202, 616

\bibitem[{{Efstathiou}(1995)}]{Efst05}
{Efstathiou}, G. 1995, \mnras, 274, L73

\bibitem[{{Gill} {et~al.}(2004){Gill}, {Knebe}, \& {Gibson}}]{amiga}
{Gill}, S.~P.~D., {Knebe}, A., \& {Gibson}, B.~K. 2004, \mnras, 351, 399

\bibitem[{{Gunn} \& {Gott}(1972)}]{gg72}
{Gunn}, J.~E. \& {Gott}, J.~R.~I. 1972, \apj, 176, 1

\bibitem[{{Hoffman} \& {Ribak}(1991)}]{HR91}
{Hoffman}, Y. \& {Ribak}, E. 1991, \apjl, 380, L5

\bibitem[{{Karachentsev}(2005)}]{kar05}
{Karachentsev}, I.~D. 2005, \aj, 129, 178

\bibitem[{{Klypin} {et~al.}(2003){Klypin}, {Hoffman}, {Kravtsov}, \&
  {Gottl{\"o}ber}}]{khkg03}
{Klypin}, A., {Hoffman}, Y., {Kravtsov}, A.~V., \& {Gottl{\"o}ber}, S. 2003,
  \apj, 596, 19

\bibitem[{{Krauss} \& {Scherrer}(2007)}]{KrasSch07}
{Krauss}, L.~M. \& {Scherrer}, R.~J. 2007, (arXiv:0704.0221)

\bibitem[{{Kravtsov} {et~al.}(2002){Kravtsov}, {Klypin}, \& {Hoffman}}]{kkh02}
{Kravtsov}, A.~V., {Klypin}, A., \& {Hoffman}, Y. 2002, \apj, 571, 563

\bibitem[{{Lahav} {et~al.}(1991){Lahav}, {Lilje}, {Primack}, \& {Rees}}]{ol91}
{Lahav}, O., {Lilje}, P.~B., {Primack}, J.~R., \& {Rees}, M.~J. 1991, \mnras,
  251, 128

%OL_23Sep07
\bibitem[{ {Lahav} \& Suto (2004){Lahav}, {Suto} }]{LS04}
{Lahav}, O., {Suto}, Y., 2004, Living Reviews in Relativity, 7, 8

\bibitem[{{Lilje} {et~al.}(1986){Lilje}, {Yahil}, \& {Jones}}]{lyj86}
{Lilje}, P.~B., {Yahil}, A., \& {Jones}, B.~J.~T. 1986, \apj, 307, 91

\bibitem[{{Loeb}(2002)}]{AL02}
{Loeb}, A. 2002, \prd, 65, 047301

\bibitem[{{Lokas} \& {Hoffman}(2001)}]{lh01}
{Lokas}, E.~L. \& {Hoffman}, Y. 2001, (astro-ph/0108283)

\bibitem[{{Maor} \& {Lahav}(2005)}]{mao05}
{Maor}, I. \& {Lahav}, O. 2005, Journal of Cosmology and Astro-Particle
  Physics, 7, 3

\bibitem[{{Martinez-Vaquero} {et~al.}(2007){Martinez-Vaquero}, {Yepes}, \&
  {Hoffman}}]{mvyh07}
{Martinez-Vaquero}, L.~A., {Yepes}, G., \& {Hoffman}, Y. 2007, \mnras, 378,
  1601

\bibitem[{{Mathis} {et~al.}(2002){Mathis}, {Lemson}, {Springel}, {Kauffmann},
  {White}, {Eldar}, \& {Dekel}}]{gif02}
{Mathis}, H., {Lemson}, G., {Springel}, V., {Kauffmann}, G., {White}, S.~D.~M.,
  {Eldar}, A., \& {Dekel}, A. 2002, \mnras, 333, 739

\bibitem[{{Nagamine} \& {Loeb}(2003)}]{NagLoeb03}
{Nagamine}, K. \& {Loeb}, A. 2003, New Astronomy, 8, 439

\bibitem[{{Peacock} \& {Schneider}(2006)}]{eso-esa-report06}
{Peacock}, J. \& {Schneider}, P. 2006, The Messenger, 125, 48

\bibitem[{{Peebles} {et~al.}(1989){Peebles}, {Melott}, {Holmes}, \&
  {Jiang}}]{PeebMel89}
{Peebles}, P.~J.~E., {Melott}, A.~L., {Holmes}, M.~R., \& {Jiang}, L.~R. 1989,
  \apj, 345, 108

\bibitem[{{Percival}(2005)}]{par05}
{Percival}, W.~J. 2005, \aap, 443, 819

\bibitem[{{Raychaudhury} \& {Lynden-Bell}(1989)}]{RayDLB89}
{Raychaudhury}, S. \& {Lynden-Bell}, D. 1989, \mnras, 240, 195

\bibitem[{{Reiprich} \& {B{\"o}hringer}(2002)}]{rei02}
{Reiprich}, T.~H. \& {B{\"o}hringer}, H. 2002, \apj, 567, 716

\bibitem[{{Spergel} {et~al.}(2006){Spergel}, {Bean}, {Dor{\'e}}, {Nolta},
  {Bennett}, {Dunkley}, {Hinshaw}, {Jarosik}, {Komatsu}, {Page}, {Peiris},
  {Verde}, {Halpern}, {Hill}, {Kogut}, {Limon}, {Meyer}, {Odegard}, {Tucker},
  {Weiland}, {Wollack}, \& {Wright}}]{wmap3}
{Spergel}, D.~N., {Bean}, R., {Dor{\'e}}, O., {Nolta}, M.~R., {Bennett}, C.~L.,
  {Dunkley}, J., {Hinshaw}, G., {Jarosik}, N., {Komatsu}, E., {Page}, L.,
  {Peiris}, H.~V., {Verde}, L., {Halpern}, M., {Hill}, R.~S., {Kogut}, A.,
  {Limon}, M., {Meyer}, S.~S., {Odegard}, N., {Tucker}, G.~S., {Weiland},
  J.~L., {Wollack}, E., \& {Wright}, E.~L. 2006, ArXiv Astrophysics e-prints

\bibitem[{{Springel}(2005)}]{gadget2}
{Springel}, V. 2005, \mnras, 364, 1105

\bibitem[{{Strauss} \& {Willick}(1995)}]{sw95}
{Strauss}, M.~A. \& {Willick}, J.~A. 1995, \physrep, 261, 271

\bibitem[{{Tegmark} \& {Rees}(1998)}]{TegRees98}
{Tegmark}, M. \& {Rees}, M.~J. 1998, \apj, 499, 526

\bibitem[{{Tonry} {et~al.}(2001){Tonry}, {Dressler}, {Blakeslee}, {Ajhar},
  {Fletcher}, {Luppino}, {Metzger}, \& {Moore}}]{sbf01}
{Tonry}, J.~L., {Dressler}, A., {Blakeslee}, J.~P., {Ajhar}, E.~A., {Fletcher},
  A.~B., {Luppino}, G.~A., {Metzger}, M.~R., \& {Moore}, C.~B. 2001, \apj, 546,
  681

\bibitem[{{Tully} \& {Fisher}(1987)}]{tf-atlas87}
{Tully}, R.~B. \& {Fisher}, J.~R. 1987, {Nearby galaxies Atlas} (Cambridge:
  University Press, 1987)

\bibitem[{{van den Bergh}(1999)}]{vdberg99}
{van den Bergh}, S. 1999, \aapr, 9, 273

\bibitem[{{Wang} \& {Steinhardt}(1998)}]{ws98}
{Wang}, L. \& {Steinhardt}, P.~J. 1998, \apj, 508, 483

\bibitem[{{Weinberg}(1987)}]{Wein97}
{Weinberg}, S. 1987, Physical Review Letters, 59, 2607

\bibitem[{{Weinberg}(2005)}]{Wein05}
---. 2005, arXiv:hep-th/0511037

\bibitem[{{Willick} {et~al.}(1997){Willick}, {Courteau}, {Faber}, {Burstein},
  {Dekel}, \& {Strauss}}]{mark3}
{Willick}, J.~A., {Courteau}, S., {Faber}, S.~M., {Burstein}, D., {Dekel}, A.,
  \& {Strauss}, M.~A. 1997, \apjs, 109, 333

\bibitem[{{Zaroubi} {et~al.}(1999){Zaroubi}, {Hoffman}, \& {Dekel}}]{zhd99}
{Zaroubi}, S., {Hoffman}, Y., \& {Dekel}, A. 1999, \apj, 520, 413

%\end{thebibliography}

\endrefs

\end{document}